\documentclass[journal]{IEEEtran}
\usepackage{graphicx}

\begin{document}
\title{Realization and Test of the Engineering Prototype of the CALICE
  Tile Hadron Calorimeter}
\author{Mark Terwort on behalf of the
   CALICE collaboration
%\thanks{Manuscript received November 12, 2010.}
\thanks{M.~Terwort is with DESY, Hamburg, Germany (e-mail:
  mark.terwort@desy.de).}
}

\maketitle
\pagestyle{empty}
\thispagestyle{empty}

\begin{abstract}
  The CALICE collaboration is currently developing an engineering
  prototype of an analog hadron calorimeter (AHCAL) for a future
  linear collider (LC) detector. It is based on scintillating tiles
  that are individually read out by silicon photomultipliers (SiPMs).
  The prototype will contain about 2500 detector channels, which
  corresponds to one calorimeter layer and aims at demonstrating the
  feasibility of building a detector with fully integrated front-end
  electronics. The concept and engineering status of the prototype, as
  well as results from the DESY test setups are reported here.
\end{abstract}

\section{Introduction}

\IEEEPARstart{A}{} new engineering prototype~\cite{Reinecke} of an
analog hadron calorimeter for a future experiment at an
electron-positron collider is currently under development by the
CALICE collaboration~\cite{CALICE}. The aim is to build a detector
with very high granularity to measure the details of hadron showers
and finally separate neutral and charged particles inside a jet.
Combining the energy information with the tracking information
improves the energy resolution. This concept is known as {\it particle
  flow} and has been tested with testbeam data taken with the CALICE
AHCAL physics prototype~\cite{Oleg}. The development of the
engineering prototype aims at the full integration of the read out
electronics into the active layers of the calorimeter to minimize dead
zones. It is based on scintillating tiles which are read out by
silicon photomultipliers (SiPMs). The full prototype will contain 2500
detector channels (one layer) and takes into account all design
aspects that are demanded by the intended operation at a LC.

A first subunit (HCAL Base Unit, HBU) with 144 detector channels has
already been produced with a size of $36\times 36$\,cm$^2$, including
the scintillating tiles, four front-end low power dissipation SPIROC
ASICs~\cite{ASICs}, the light calibration and gain-monitoring system
and the detector/DAQ interface boards that are used for power supply
as well as slow control programming. The power-supply module allows
for switching off individual detector components within the LC
bunch-train structure ({\it power pulsing}). A new data-aquisition
(DAQ) has been set up, including a graphical user interface based on
Labview 8.6 for comfortable test operation, while the development and
testing of the final DAQ system~\cite{Reinecke} for the full prototype
is still ongoing.

Currently, these HBU sub-components are redesigned in order to
optimize the spatial dimensions as well as the performance of the
ASICs, tiles, calibration systems and detector/DAQ interface modules
to match the requirements for the integration of a full AHCAL layer.
In this report the concept of the design and results from test
measurements are presented.

\section{Concept and Design of the Engineering Prototype}

\begin{figure}%[!t]
\centering
\includegraphics[width=3.5in]{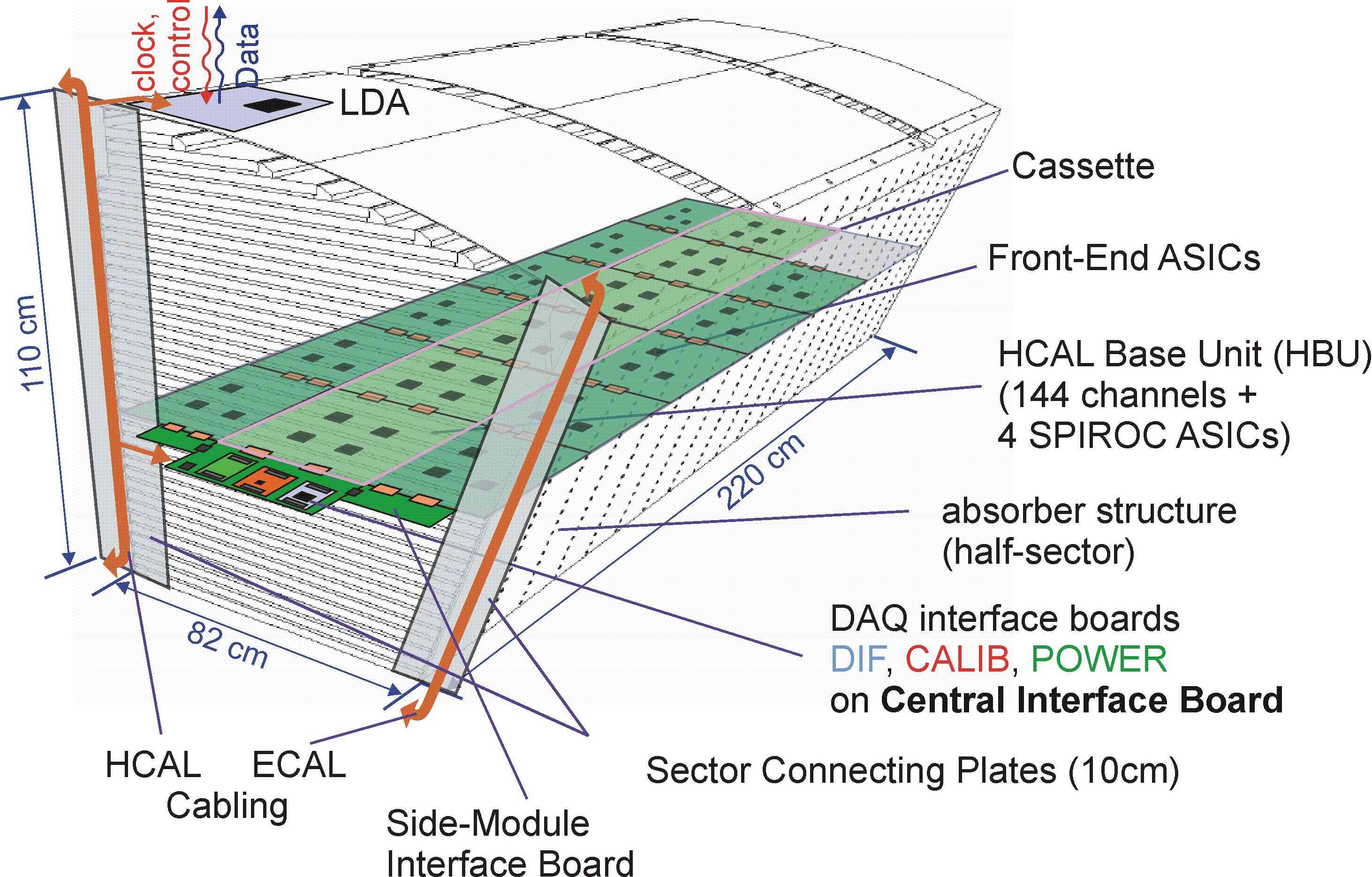}
\caption{1/16 HCAL half-barrel with 48 layers containing three slabs each.}
\label{segment}
\end{figure}

The barrel of the AHCAL has a cylindrical structure and will be placed
outside the electromagnetic calorimeter, while it is surrounded by the
magnet. The inner and outer radius is 1.8m and 2.8m, respectively. The
cylindrical structure is divided into 16 segments with 48 detector
layers each. One layer consists of the tiles, the embedded front-end
electronics and a $\sim 16$\,mm thick stainless steel or $\sim 10$\,mm
thick tungsten absorber plate. The typical size of a segment's layer
is $1\times 2.2$\,m$^2$. With 2500 channels per layer the total number
of channels of the AHCAL barrel adds up to about 3.9 million.

\begin{figure}%[!t]
\centering
\includegraphics[width=2.5in]{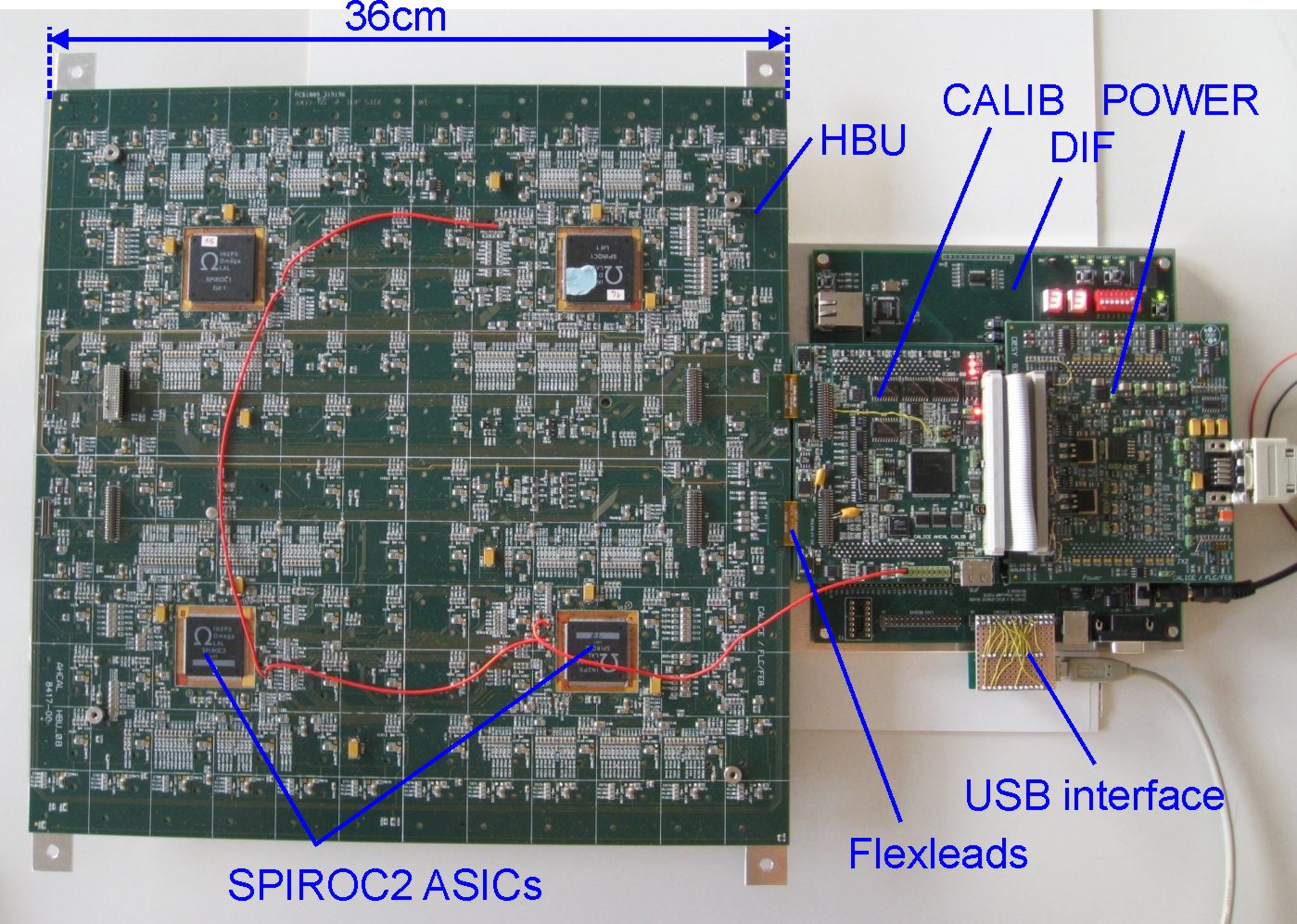}
\caption{Current setup of an HBU as used in the DESY test measurements.}
\label{HBU}
\end{figure}

\begin{figure}%[!t]
\centering
\includegraphics[width=2.5in]{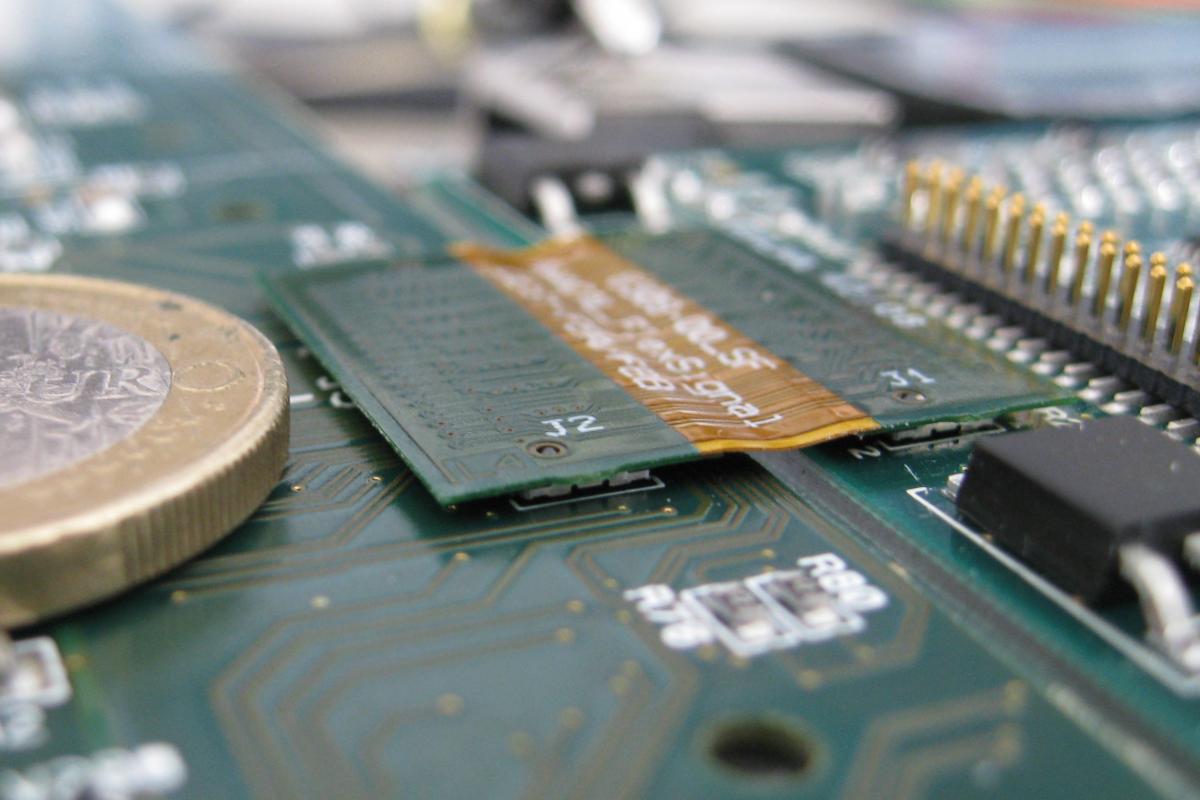}
\caption{HBU interconnection via flexleads and ultra-thin connectors.}
\label{Flexfoil}
\end{figure}

Fig.~\ref{segment} depicts the design of a single segment, showing the
implementation of a specific active layer consisting of three parallel
slabs. Each slab consists of six HBUs and the middle slab is connected
to the DAQ via the Central Interface Board (CIB). The side slabs are
in turn connected to the CIB via the Side Interface Boards (SIBs). The
first HBU module, along with the interface modules, is shown in
Fig.~\ref{HBU}, as it is used in the DESY test setups. In the final
design the HBUs are interconnected by flexleads and ultra-thin
connectors with a stacking height of 0.8\,mm (Fig.~\ref{Flexfoil}),
which are also used to connect the HBUs to the CIB.

One of the main new features of the embedded electronics is the
on-detector zero suppression. The measurements in the testbeam will
show how well this can be controlled with realistic spreads of
scintillator light yields and photo-sensor gains as well as changes in
environmental conditions. The chips are designed to operate with
pulsed power supply for minimized heat dissipation. Establishing this
operating mode under beam conditions is a major step towards
establishing the integration concept for a highly granular
calorimeter.

\subsection{Tiles and ASICs}

\begin{figure}%[!t]
\centering
\includegraphics[width=2.5in]{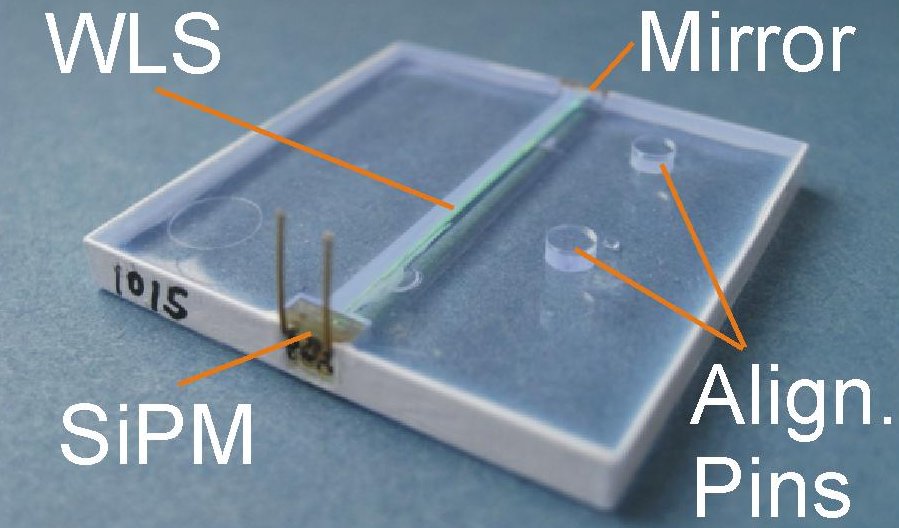}
\caption{Scintillating tile with embedded wavelength shifting fiber,
  SiPM, mirror and alignment pins.}
\label{Tile}
\end{figure}

The signal that is detected by the SiPMs is produced by scintillating
tiles with a size of $3\times 3\times 0.3$\,cm$^3$, as shown in
Fig.~\ref{Tile}. The new design differs from the design used in the
physics prototype~\cite{PPT} and includes a straight wavelength
shifting fiber coupled to a SiPM with a size of 1.27\,mm$^2$ on one
side and to a mirror on the other side. The SiPM comprises 796 pixels
with a gain of $\sim 10^6$. Two alignment pins are used to connect the
tiles to the HBU's printed circuit board (PCB) by plugging them into
holes in the PCB. The nominal tile distance is 100\,$\mu$m.

\begin{figure}%[!t]
\centering
\includegraphics[width=2.5in]{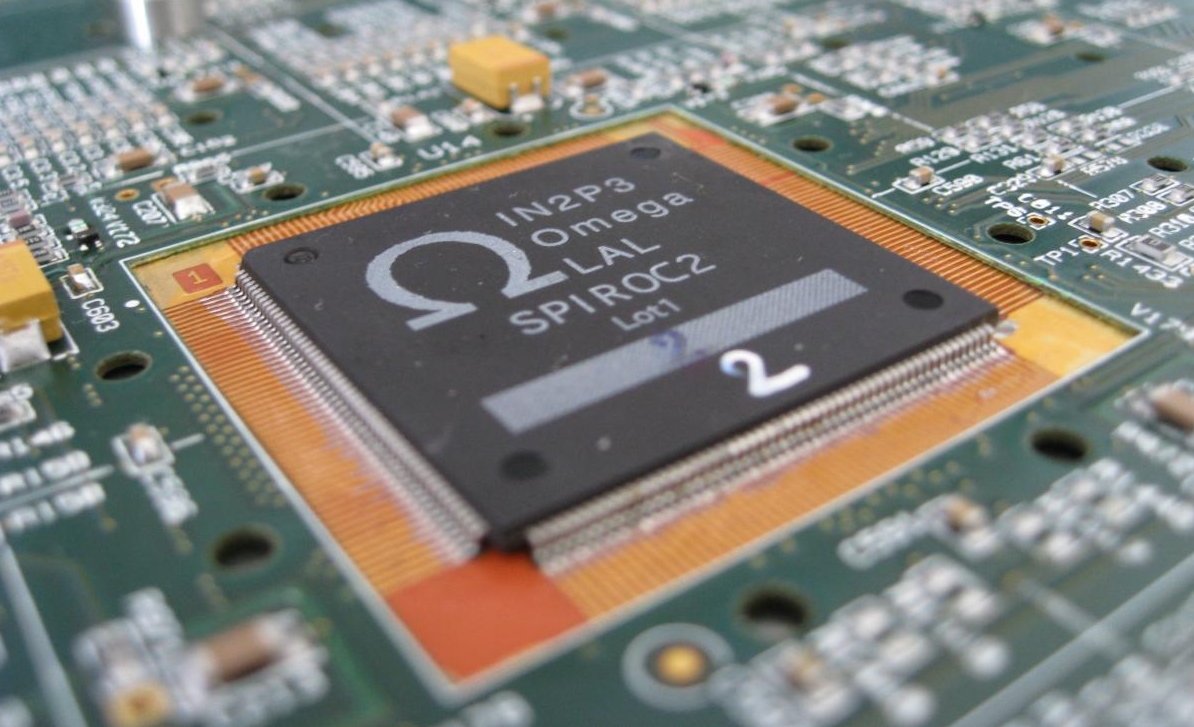}
\caption{Integration of the SPIROC ASIC into the PCB.}
\label{SPIROC_sink}
\end{figure}

For each HBU the analog signals from the SiPMs are read out by four
36-channel ASICs equipped with 5\,V DACs for a channel-wise bias
voltage adjustment. They provide two gain modes, which leads to a
dynamic range of 1 to 2000 photo electrons. The foreseen power
consumption amounts to 25\,$\mu$W per channel for the final LC
operation. The main new features of the ASICs compared to the physics
prototype are the integration of the digitization step (12-bit ADC and
12-bit TDC for charge and time measurements) and the self-triggering
capability with an adjustable threshold. To reduce the height of the
active layers the ASICs are lowered into the PCB by $\sim
500$\,$\mu$m. This leads to a total reduction of the AHCAL diameter of
48\,mm. A picture of an ASIC as it is embedded into the PCB is shown
in Fig.~\ref{SPIROC_sink}.

\subsection{Detector/DAQ Interface}

\begin{figure}[!b]
\centering
\includegraphics[width=3.5in]{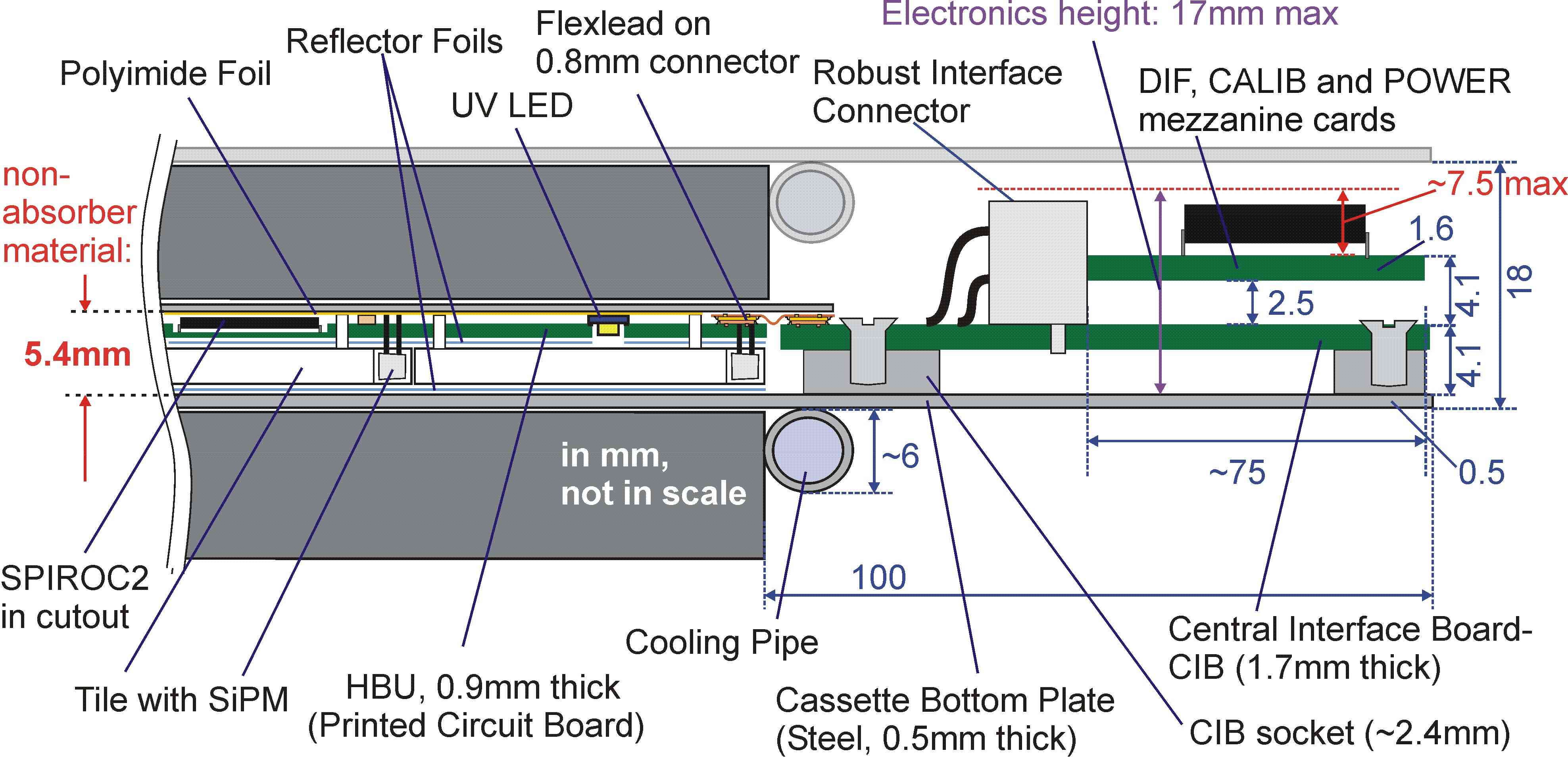}
\caption{Cross section of one HCAL layer, including the absorbers,
  tiles, SiPMs, PCBs, ASICs and the detector/DAQ interface modules.
  All dimensions are given in mm.}
\label{HCAL_cross_endface}
\end{figure}

\begin{figure}%[!t]
\centering
\includegraphics[width=3.5in]{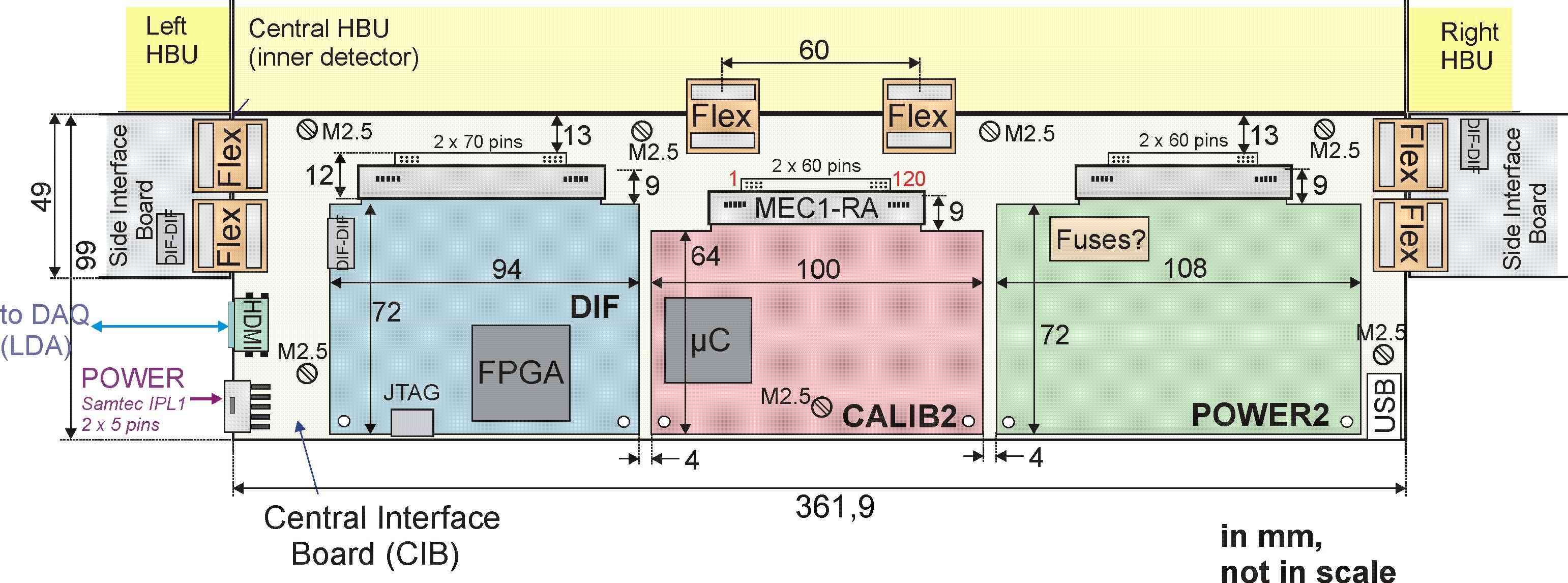}
\caption{Schematics of the CIB hosting the DIF, CALIB and POWER
  modules.}
\label{CIB}
\end{figure}

Fig.~\ref{HCAL_cross_endface} shows the cross section of one AHCAL
layer including the dimensions of the single components. The active
layers, including the tiles, SiPMs, PCBs and ASICs, are shown as they
are placed between two layers of absorber material and connected to
the CIB. The total height of the detector/DAQ interface modules hosted
by the CIB has to be very small ($\sim$~18\,mm in case of a tungsten
absorber) in order to fit between two layers. The details of the
arrangement of the interface modules are shown in Fig.~\ref{CIB}. The
connections of the CIB to the SIBs and the HBUs via thin flexlead
connectors are visible.  The Detector Interface (DIF) serves as the
interface between the inner-detector ASICs and the DAQ. The full
operation via a USB-debug interface and Labview is possible and has
been used in all measurements performed so far. The CALIB module
controls the UV LED system, while the POWER module generates all
detector supply voltages.  It also enables the LC power pulsing.

\subsection{Light Calibration System}

\begin{figure}[!b]
\centering
\includegraphics[width=2.5in]{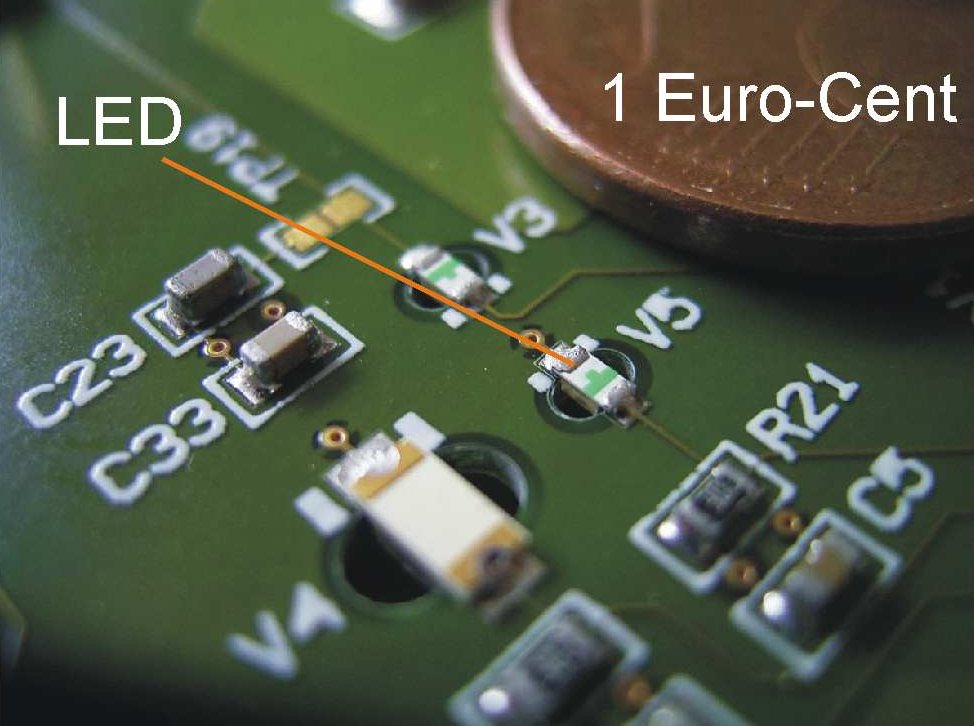}
\caption{Light calibration system with one integrated LED per tile.}
\label{LED_Wuppertal}
\end{figure}

Since the SiPM response shows a strong dependence on the temperature
and bias voltage and saturates due to the limited number of pixels, a
gain-calibration and saturation-monitoring system with a high dynamic
range is needed. In the calibration mode of the ASICs a very low light
intensity is needed to measure the gain as the distance between the
peaks in a single-pixel spectrum, while at high light intensities
(corresponding to $\sim$100 minimum-ionizing particles (MIPs)) the
SiPM shows saturation behaviour. Currently there are two concepts
under investigation:
\begin{itemize}
\item One LED per tile that is integrated into the detector gap
  (Fig.~\ref{LED_Wuppertal}). This system is used in the HBUs in the
  DESY test setups.
\item One strong LED outside the detector, while the light is
  distributed via notched fibers (see~\cite{Ivo}).
\end{itemize}

\begin{figure}%[!t]
\centering
\includegraphics[width=3.5in]{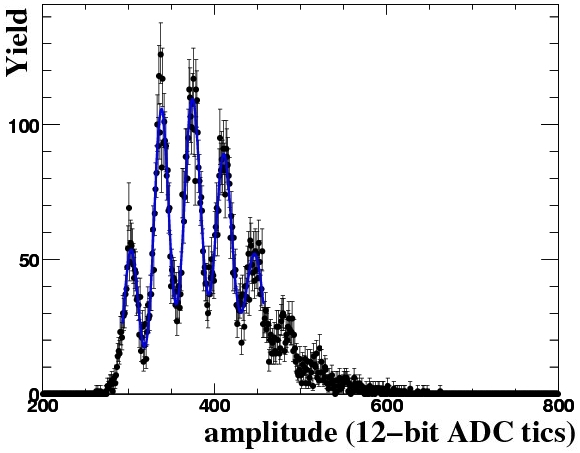}
\caption{Typical single-pixel spectrum as measured with the HBU in the
  DESY test setup. The distance between the peaks is used for the gain
  calibration.}
\label{SPS}
\end{figure}

Both options have been successfully tested on the DESY test setups in
the laboratory and under testbeam conditions. A typical singel-pixel
spectrum as measured with the integrated calibration system of the
test HBUs together with a fit to the gaussian peaks is shown in
Fig.~\ref{SPS}. The measured cross-talk is purely optical and is of
the order of 2.5\%. The dynamic range of the system redesigned for the
construction of the engineering prototype is currently under
investigation. The channel uniformity is also an open issue, since for
the first LED system the individual LEDs have a large spread of the
emitted light intensity, while for the second system the light
coupling from the fiber to the tile and its mechanical integration in
a full prototype is unsolved.

\section{Measurements and Results}

\begin{figure}[!b]
\centering
\includegraphics[width=2.5in]{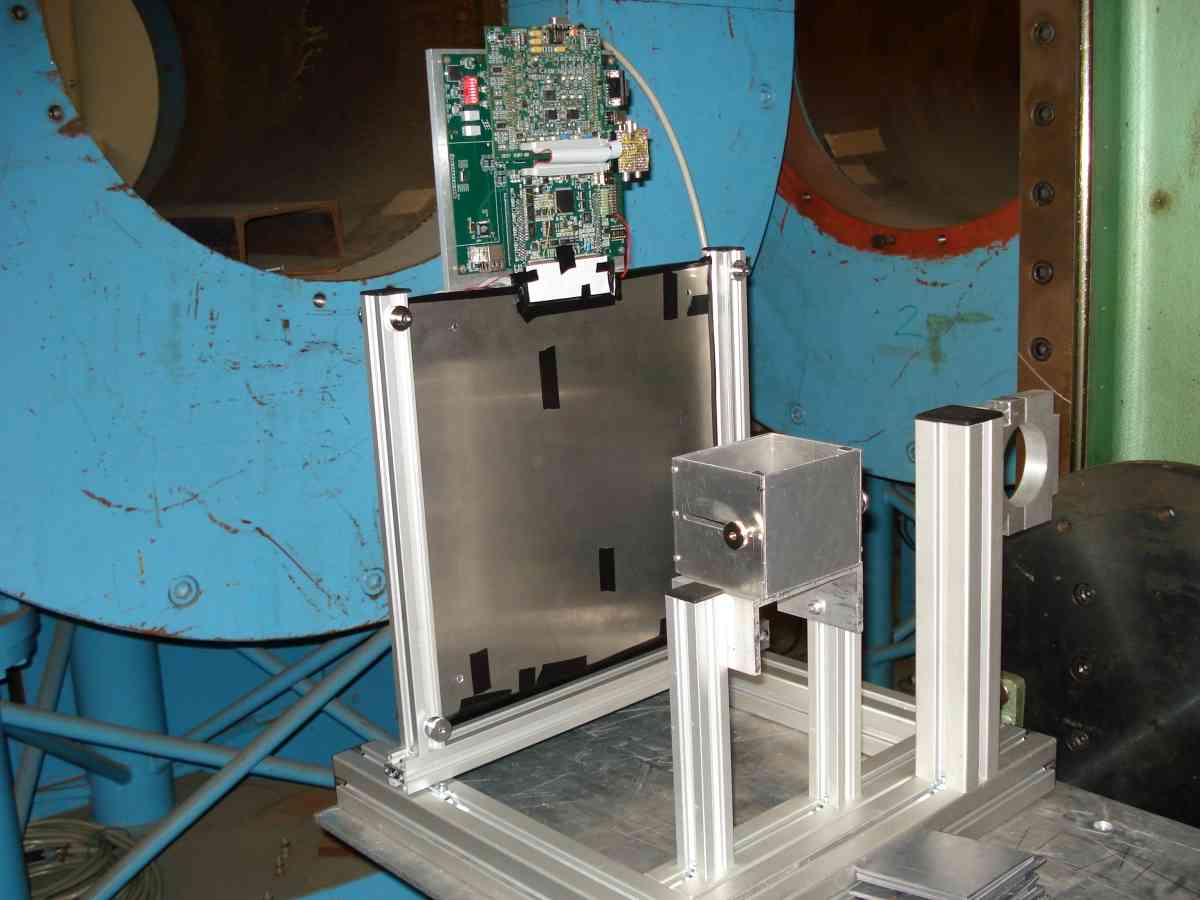}
\caption{The setup of the light-tight HBU cassette at the DESY
  testbeam is shown. On top of the cassette the CIB hosting the
  detector/DAQ interface modules is visible. The whole setup is
  mounted on a movable stage in order to be able to scan all
  channels.}
\label{testbeam}
\end{figure}

\begin{figure}%[!t]
\centering
\includegraphics[width=3.5in]{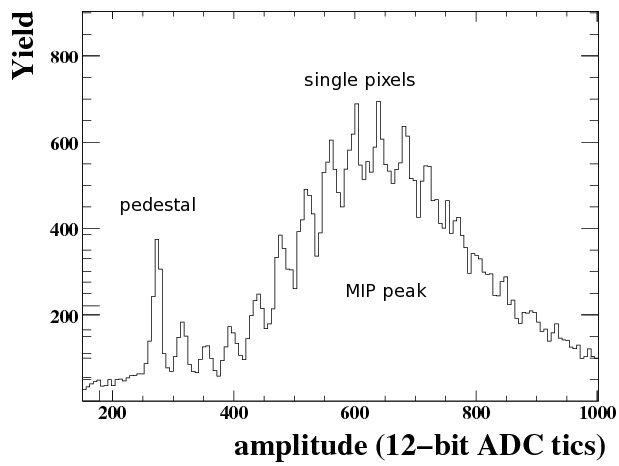}
\caption{Measurement result of a typical MIP spectrum obtained with
  the HBU and the DAQ interface modules using the newly developed
  Labview/USB DAQ in the 6\,GeV DESY electron testbeam.}
\label{MIP}
\end{figure}

The main task of the current characterization is to prove the
suitability of the realized detector-module concept for the
larger-scale prototype with 2500 channels and the final length of
2.2\,m. Two setups as shown in Fig.~\ref{HBU} are in operation, one in
the DESY 6\,GeV electron testbeam facility (the 2-6\,GeV electrons
that have been used are MIPs in the scintillating tiles) and the
second in a laboratory environment. A picture of the current testbeam
setup is shown in Fig.~\ref{testbeam}. The HBU is enclosed into a
light-tight aluminum cassette and is mounted on top of a movable stage
in order to be able to scan all channels. The external beam trigger
consists of two 10\,cm long scintillator counters, which are placed in
front of the module and are required to give coincident signals. The
trigger starts the data taking on the board as far as a spill signal
from the machine is present. For the results presented in the
following the high gain mode of the ASIC is used with a feedback
capacitance of 100\,fF to couple the SiPM signal into the front-end
electronics and a shaping time of 50\,ns.

\begin{figure}%[!t]
\centering
\includegraphics[width=3.5in]{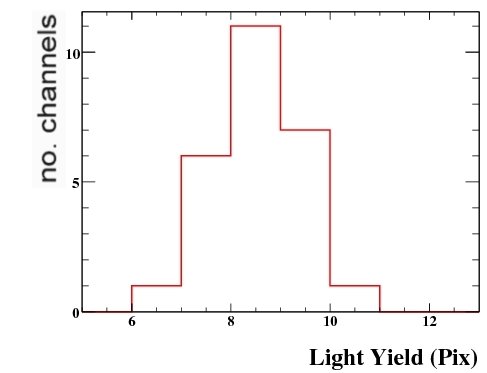}
\caption{Light yield as measured in the DESY testbeam facility for
  multiple channels of the HBU.}
\label{LY}
\end{figure}

After the investigation of the fundamental properties like noise
behaviour and signal delays~\cite{Riccardo}, measurements in the
laboratory using the LED calibration system and a charge injection
setup as well as testbeam measurements have been performed to
investigate the uniformity of the SiPM response as well as the
behaviour of the tiles for multiple channels. Fig.~\ref{MIP} shows a
typical MIP spectrum measured with the electron testbeam. It can be
seen that single pixel peaks are clearly distinguishable for more than
10 peaks. The first peak is the pedestal peak and the maximum of the
spectrum is at 9 pixels. The distribution of the light yields, defined
as the most probable number of active pixels for a MIP signal, is
plotted in Fig.~\ref{LY}.

In the following some results of the investigation of the auto-trigger
are presented~\cite{Jeremy}.

\subsection{The Auto-Trigger}

\begin{figure}%[!t]
\centering
\includegraphics[width=3.5in]{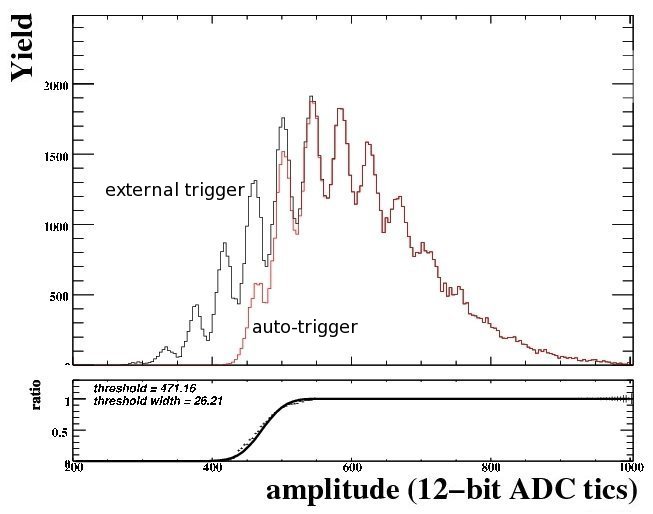}
\caption{Comparison of a single-pixel spectrum produced with LED light
  for external- and auto-triggering with a given threshold.}
\label{AT_LED_MIP}
\end{figure}

For running in a LC environment the ASICs need the capability of
self-triggering (auto-triggering), since the LC does not provide a
central global trigger signal. Therefore the analog amplitude from the
SiPM is shaped with a fast shaper (25\,ns) and compared to a
predefined adjustable threshold by a discriminator to make the trigger
decision. Fig.~\ref{AT_LED_MIP} shows two single-pixel spectra
measured with LED light and external trigger (black histogram) and
auto-trigger (red histogram), respectively. The ratio of the red
histogram divided by the black histogram is also shown and gives an
impression of the width of the trigger turn-on curve. After the
turn-on the trigger efficiency is 100\%. The predefined trigger
threshold used in this plot is of the order of 0.5\,pC.

\begin{figure}%[!t]
\centering
\includegraphics[width=3.5in]{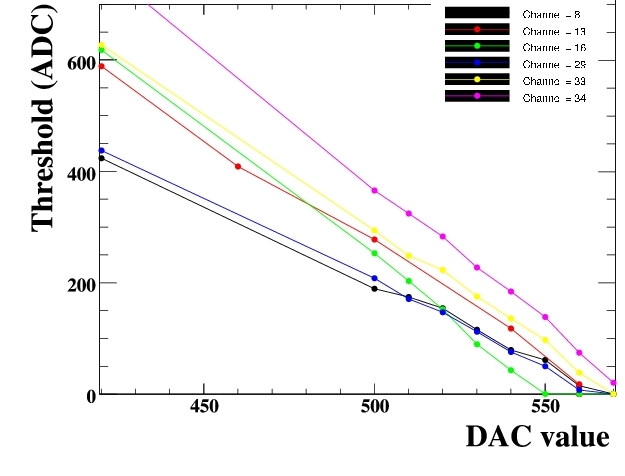}
\caption{Measured auto-trigger threshold as a function of the
  adjustable 10-bit DAC threshold. The plot shows several channels
  investigated with the DESY testbeam.}
\label{BeamDACvsThreshold}
\end{figure}

The behaviour of the measured threshold as a function of the
predefined 10-bit DAC threshold has been investigated in charge
injection as well as in testbeam measurements.
Fig.~\ref{BeamDACvsThreshold} shows a linear behaviour for the
testbeam measurement, but a significant spread among several channels
in the measured threshold for a given DAC setting. This shows the
necessity of having the possibility to adjust the auto-trigger
threshold channel-wise.

\begin{figure}%[!t]
\centering
\includegraphics[width=3.5in]{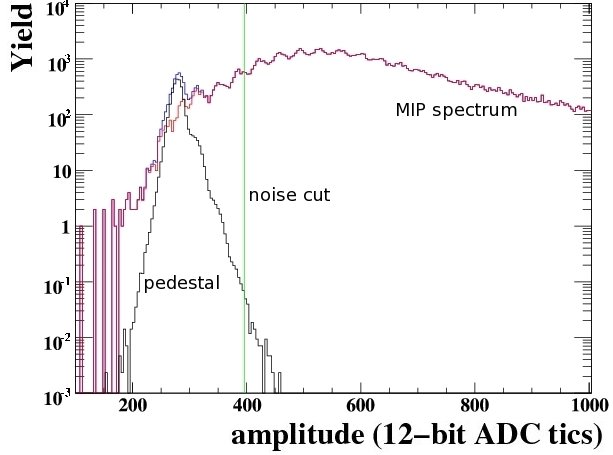}
\caption{Calculated auto-trigger threshold for having a noise to
  signal ratio of smaller than 10$^{-4}$. An independent pedestal
  measurement is compared to a MIP spectrum.}
\label{Pedestal_cut}
\end{figure}

\begin{figure}%[!t]
\centering
\includegraphics[width=3.5in]{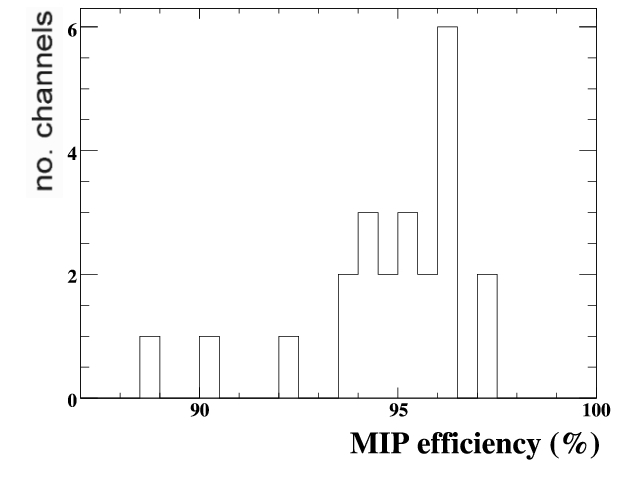}
\caption{Distribution of the MIP detection efficiency for a calculated
  auto-trigger threshold (noise to signal ratio smaller than
  10$^{-4}$).}
\label{MIP_distribution}
\end{figure}

The threshold of the auto-trigger will be adjusted in order to
minimize the noise hits and simultaneously maximize the efficiency for
measuring a MIP. To investigate the distribution of the MIP
efficiencies, the position of the pedestal cut has been defined such
that the number of noise hits divided by the number of MIP events is
smaller than 10$^{-4}$. To calculate this, an independent noise
measurement has been compared to the testbeam measurement of MIP
events. Fig.~\ref{Pedestal_cut} shows the MIP distribution (red
histogram) for a given channel together with the pedestal measurement
of the same channel (black histogram). The green line indicates the
calculated auto-trigger threshold. With the measurement shown in
Fig.~\ref{BeamDACvsThreshold} this threshold can be translated into a
trigger threshold DAC setting for the slow control data. It turns out
that a MIP efficiency of around 95\% is achieved with this method. The
distribution for multiple channels is shown in
Fig.~\ref{MIP_distribution}.

\section{Conclusions and Outlook}

The CALICE collaboration is actually developing an engineering
prototype for the AHCAL technology option for a possible LC
experiment. It is foreseen to build a complete calorimeter layer with
fully integrated front-end electronics to demonstrate the feasibility
and scalability of the concepts. This requires the redesign of all
involved components, including ASICs, scintillating tiles, HBUs, LED
calibration systems and the detector/DAQ interface modules. The tests
and measurements with charge injection, LED light and testbeams that
have been performed so far, show promising results in terms of
functionality and performance and lead to an increased understanding
of the system. Among them are tests of the collective behaviour of
many channels in terms of uniformity and interdependences as well as
investigations of the behaviour of the auto-trigger.

Two of the most important next steps to be taken are tests of the
power pulsing capabilities in a testbeam environment and the
integration of the next generation DAQ. Additionally, there will be
detailed studies of the TDC behaviour as well as further tests of the
single next-generation components, like ASICs and tiles, to reach the
performance goals required for a LC experiment.

\section{Acknowledgment}

The author gratefully thanks Erika Garutti, Peter G\"ottlicher,
Mathias Reinecke, J\'er\'emy Rou\"en\'e, Julian Sauer and Felix Sefkow
for very useful discussions and valuable contributions to the results
presented here. This work is supported by the Commission of the
European Communities under the 6th Framework Programme ``Structuring
the European Research Area'', contract number RII3-026126.


\begin{thebibliography}{1}

\bibitem{Reinecke} M.~Reinecke et al., \emph{Integration Prototype of
    the CALICE Tile Hadron Calorimeter for the International Linear
    Collider}, Proc.~2008 IEEE Nuclear Science Symposium (NSS08),
  NSSMIC.2008.4774800; \mbox{T.\ Buanes} et al., \emph{The CALICE
    Hadron Scintillator Tile Calorimeter Prototype}, Proc. of Techn.
  and Instr. in Part. Physics (TIPP), 2009, to be published in Nucl.
  Instr. and Meth. A; P.~G\"ottlicher for the CALICE collaboration,
  \emph{First Results of the Engineering Prototype of the CALICE Tile
    Hadron Calorimeter}, Proc.~2009 IEEE Nuclear Science Symposium
  (NSS09), NSSMIC.2009.5402334.

\bibitem{CALICE} CALICE home page: \\
  https://twiki.cern.ch/twiki/bin/view/CALICE/CaliceCollaboration

\bibitem{Oleg} O.~Markin for the CALICE collaboration,
  \emph{PandoraPFA Tests using Overlaid Charged Pion Test Beam Data},
  to be published in CALOR2010 proceedings.

\bibitem{ASICs} L.~Raux et al., \emph{SPIROC Measurement: Silicon
    Photomultiplier Integrated Readout Chips for ILC}, Proc.~2008 IEEE
  Nuclear Science Symposium (NSS08), NSSMIC.2009.5401891; R.~Fabbri,
  \mbox{B.~Lutz} and W.~Shen, \emph{Overview of Studies on the
    SPIROC Chip Characterisation}, arXiv:0911.1566 and
  EUDET-Report-2009-05, October 2009.

\bibitem{PPT} C.~Adloff et al., \emph{Construction and commissioning
    of the CALICE analog hadron calorimeter prototype}, JINST 5 (2010)
  P05004, arXiv:1003.2662

\bibitem{Ivo} I.~Pol\'ak, \emph{An LED calibration system for the CALICE
    HCAL}, these proceedings, 2010.

\bibitem{Riccardo} R.~Fabbri for the CALICE collaboration,
  \emph{CALICE Second Generation AHCAL Developments}, Proc. 2010 LCWS, arXiv:1007.2358.

\bibitem{Jeremy} J.~Rou\"en\'e, \emph{Analysis of the autotrigger of the
    read out chip of the front-end electronics for the HCAL of the
    ILC}, summer student report, 2010.

\end{thebibliography}
\end{document}